\documentstyle[12pt,titlepage]{article}
\newcommand{\ba}{\begin{array}}
\newcommand{\ea}{\end{array}}
\newcommand{\bd}{\begin{displaymath}}
\newcommand{\ed}{\end{displaymath}}
\newcommand{\be}{\begin{equation}}
\newcommand{\ee}{\end{equation}}
\newcommand{\bea}{\begin{eqnarray}}
\newcommand{\eea}{\end{eqnarray}}

\begin{document}
\setcounter{page}{0}
\thispagestyle{empty}

\begin{flushright} {UCRHEP-T163} \\
July 1996 \\
\end{flushright}

\begin{center} {\Large\bf SUPERSYMMETRY AND THE SPONTANEOUS BREAKING OF A U(1)
GAUGE FACTOR\\[0.5truein]} {E. Keith$^1$, Ernest Ma$^2$ 
and Biswarup Mukhopadhyaya$^3$\\}
Department of Physics\\ University of California\\ Riverside, CA 92521, USA
\end{center}

\vskip 25pt

\begin{center} {\large\bf ABSTRACT}
\end{center}

We investigate a supersymmetric theory with an extra U(1) gauge symmetry 
surviving down to
low energies. The extra U(1) is assumed to originate from an E$_6$ grand 
unified theory
(GUT). We show that if one assumes universal soft supersymmetry breaking 
scalar masses at
the GUT scale, and requires the mass of the additional U(1) gauge boson 
to satisfy
phenomenological bounds, then the conditions for electroweak symmetry 
breaking will provide
stringent restrictions on the allowed parameter space of such a theory. 
We also determine
the masses of standard as well as exotic sfermions and find that 
it is possible for the latter to be lighter
than the former.  An interesting specific observation is 
that a light stop is
difficult to accommodate with an extra U(1) gauge symmetry.

\vskip 40pt

\footnotesize
\noindent
$^{1}$ E-mail : evan@citrus.ucr.edu \\
$^{2}$ E-mail : ernestma@citrus.ucr.edu\\ $^{3}$ Permanent address: 
Mehta Research
Institute, 10 Kasturba Gandhi Marg, Allahabad - 211 002, India. E-mail: 
biswarup@mri.ernet.in \normalsize 

\newpage
\textheight=8.9in

\section{INTRODUCTION}

Supersymmetry (SUSY) is perhaps the most widely pursued option in the quest 
for physics
beyond the standard model \cite{susy}. Though the efforts in this direction 
are woven largely around
the minimal supersymmetric standard model (MSSM), the importance of 
extending beyond the
minimal case should also be recognized. There are at least two main reasons: 
(1) The mechanism for
SUSY breaking is not yet understood, and a viable mechanism (although 
not necessarily the
only one) consists in breaking SUSY softly at a very high scale, for 
example, through
gravitational interactions \cite{sugra}; (2) Recent measurements of the three 
coupling constants of the
standard model have strongly suggested that they tend to unify at 
an energy scale of
$10^{16}$ GeV in a supersymmetric scenario \cite{uni}. Thus there is enough 
motivation to think that
if Nature is indeed supersymmetric, then SUSY is perhaps embedded 
in a framework of higher
symmetry which, manifested at a high scale, ultimately leads to softly 
broken SUSY and all
its desirable features at a scale of 1 TeV or below. 

A class of candidate theories in this context consists of the E$_6$ 
grand unified theories
(GUTs) which were originally motivated by superstrings. One generic 
feature of such theories
is that they predict an extra U(1) symmetry surviving down to low 
energies \cite{[hewriz]}. Consequently, as we stand close to the 
exploration of
phenomenology at the TeV scale, it is useful to know all possible 
effects of this extra
U(1) . For example, the spontaneous breaking of this U(1) leads to 
an additional neutral
gauge boson whose phenomenological implications are under close 
scrutiny today. It has also
been suggested
\cite{[N1],[N2]} that such a scenario can lead to light singlet 
neutrinos which offer a
possible reconciliation of the recent LSND data on neutrino 
oscillations \cite{lsnd} with the solar and
atmospheric neutrino puzzles. 

In this paper, we consider a general supersymmetric E$_6$ scenario, 
and discuss how the
breakdown of the extra U(1) at the scale of about 1 TeV affects the 
sfermion masses in a
SUSY model. With a universal scalar mass of size $m_0$ at the GUT 
scale, all the scalar
masses at low energy are determined by the gaugino contributions to 
the running masses, as
well as by the D-term contributions which arise from the reduction 
in rank of the gauge
group at the SUSY breaking scale. In particular, the large vacuum 
expectation value (VEV)
of an isosinglet scalar field, which in turn gives mass to the 
additional gauge boson,
leads to D-terms of substantial magnitude. We shall show that this 
affects the Higgs mass
parameters in such a way that the model becomes quite restricted from 
the standpoint of
electroweak symmetry breaking. In addition, it predicts certain 
qualitative features among
the sfermion masses. The combinations of parameters which in general 
fulfill all
constraints normally lead to sfermion masses beyond the reach of LEP-2, 
and perhaps
beyond that of the Tevatron as well. Exotic sfermions lighter than ordinary 
ones can be
envisioned in such a scenario. Also, the large D-terms make it difficult to 
accommodate a
light stop, which is a rather interesting feature of such theories. 

In Section 2 we discuss the salient features of a SUSY model with an extra 
U(1) from E$_6$.
We also list there the general expressions for different scalar masses in 
this framework.
In Section 3 the constraints that follow on model parameters are discussed 
in the light of
these expressions. Section 4 contains some sample numerical results. There 
we also point
out the improbability of a light stop in such a scenario. We conclude in 
Section 5. The
appendix contains some general observations on the feasibility of breaking 
an extra U(1)
symmetry while preserving SUSY.

\section{GENERAL MODEL AND SCALAR MASSES} 

As has been previously stated, we consider a supersymmetric E$_6$ grand 
unified theory. The
extra U(1)'s in E$_6$ can be described by the following decomposition: 

\begin{equation} E_6 \longrightarrow SO(10) \times U(1)_\psi \end{equation} 

and

\begin{equation} SO(10) \longrightarrow SU(5) \times U(1)_\chi\, . 
\end{equation} 

A rank-5 low-energy theory can be obtained if a linear combination of 
$U(1)_{\psi}$ and
$U(1)_{\chi}$ survives down to the TeV scale: 

\begin{equation} U(1)_{\psi} \times U(1)_{\chi} \longrightarrow U(1)_{\alpha},
\end{equation}

\noindent where the charge for the surviving extra U(1) is, in general, 

\begin{equation} Q_{\alpha} = Q_{\psi} \cos \alpha - Q_{\chi} \sin {\alpha} 
\end{equation}

The particle content we will choose to use here is three complete {\bf 27}'s 
of E$_6$ along
with an extra pair of color-singlet SU(2)$_L$ doublet fields. The different 
quantum
numbers, including the
$\psi$- and
$\chi$-charges of the various superfields are found in Table 2 of Ref. 
\cite{[hewriz]}.
However, we note that different conventions are followed by different 
authors. In one-loop
order, our choice of field content leads to a unification of the gauge 
couplings at the
same scale as in the MSSM. We also note that, due to the addition of 
exotic colored fields,
$\alpha_s$ does not run in one-loop order. Of course, one could also 
have assumed that
gauge-coupling unification comes from threshold corrections which could 
then occur at about
the string scale.

The angle $\alpha$ is unspecified in general. The most common specific case 
is where E$_6$
breaking takes place via Wilson loops, leading to the
$\eta$-model, where in our convention, $\alpha = \tan^{-1} \sqrt{3/5}$. 
Another example is
a recently proposed model, henceforth to be referred to as the $N$-model, 
where light singlet
neutrinos can be generated, which corresponds to $\alpha = -\tan^{-1} 
\sqrt{1/15}$. Details
of this latter model can be found in Ref. \cite{[N1]}.

The {\bf 27} of E$_6$ contains all the matter fields surviving down to the 
TeV scale. In
addition to the particles belonging to the MSSM, each generation also has 
the isosinglet
color-triplets
$h$ and $h^c$, and the isosinglet color-singlets $\nu^c$ and $S$, the latter 
being trivial 
under $U(1)_{\chi}$ as well. One can construct the model so as to break
$U(1)_{\alpha}$ spontaneously with the VEV of the scalar component of $S$ 
(as well as the
two SU(2) doublet Higgs VEVs):

\begin{eqnarray} \langle{\tilde S}\rangle = u\, ,\, \langle H_1 \rangle = 
v_1\, ,\, \langle
H_2 \rangle = v_2\, , \end{eqnarray}

\noindent where $H_1$ and $H_2$ are the two Higgs doublets giving masses to 
the down- and
up-type quarks respectively, and $\tan \beta = v_{2}/v_{1}$. It is the 
singlet VEV $u$ which
sets the scale of $U(1)_{\alpha}$ breaking, to be constrained by 
experimental lower bounds
on the mass of $Z'$, the additional neutral gauge boson. There is 
in general mixing between
$Z$ and $Z'$, and the physical states can be obtained by diagonalizing 
the mass matrix. In
practice, assuming $g_{\psi} = g_{\chi} = g_Y$ from a GUT hypothesis, the
$Z'$ mass is approximately given by

\begin{equation} m_{Z'}^2 = {\frac{4}{3}} g^2_Y u^2 \cos^{2} {\alpha} + 
O(v^2_1, v^2_2)\, .
\end{equation}

\noindent Note that a value of $\alpha$ very close to $\pi/2$ will lead 
to an inadmissibly
small $Z'$ mass, and is hence disallowed for $u$ in the TeV range.

Let us now consider the sfermion masses in this scenario. We assume that 
soft SUSY breaking
scalar terms are generated via gravitational interactions with a hidden 
sector at the GUT
scale $M_G$. We further make the conventional assumption that these soft 
breaking terms are
flavor blind. Consequently, at the scale $M_G$, there is a universal mass
$m_0$ for all scalar fields, a universal gaugino mass $M_i =m_{1/2}$, and 
all trilinear soft
SUSY breaking scalar terms are parameterized by a universal massive 
parameter $A_0$. At the
one-loop level, we have the gluino mass $m_{\tilde{g}} =m_{1/2}$ because
the one-loop beta function vanishes for the strong coupling. The scalar masses 
at low energy are
obtained by running the renormalization group equations (RGEs) down to 
the weak scale and
including the D-term masses arising from both $U(1)_{\alpha}$ and 
electroweak symmetry
breaking. We include F-term masses only when calculating the stop 
masses. The RGEs yield
the following expressions for the scalar masses at the scale $\mu$: 
\begin{eqnarray}
m_{H_2}^2&=&m_0^2+\sum_i C_i^{(H_2)}-{3\over 2}I\, ,\\ 
m_{\tilde{Q}_3}^2&=&m_0^2+\sum_i
C_i^{(Q_3)}-{1\over 2}I\, ,\\ m_{\tilde{t}^c}^2&=&m_0^2
+\sum_i C_i^{(t^c)}-I\, ,\\
m_{R}^2&=&m_0^2+\sum_i C_i^{(R)}\, ,
\end{eqnarray} where $R$ represents all other fields except 
those appearing in the trilinear 
superpotential term
$\lambda_t H_2 \tilde{Q}_3 \tilde{t}^c$. We have assumed that 
only the top Yukawa coupling,
which appears through
\begin{eqnarray} I={1\over 4\pi^2}\int^{t_G}_{t}\lambda_t^2 \left(
A_t^2+m_{H_2}^2+m_{\tilde{Q}_3}^2+m_{\tilde{t}^c}^2\right) dt 
\end{eqnarray} has a
significant effect on the running of the scalar masses. In the 
above equation, the gaugino
loop contribution for a scalar field $F$ is given as \cite{ram}
\begin{eqnarray}
C_i^{(F)}=c_2(F_i)
\cdot {1\over 2\pi^2}\int^{t_G}_{t} g_i^2 M_i^2dt \, ,
\end{eqnarray} with $t= \ln{\left( \mu/{\rm GeV}\right) }$ and 
$t_G= \ln{\left( \mu_G/{\rm
GeV}\right) }$. Both of these integrals can be solved analytically
at the one-loop level \cite{[rgesoln]}.

At the weak scale and at the scale $u$, D-term contributions to 
these masses must be added
to the previous expressions. Although in our numerical calculation 
we have included
$O(v_1^2,v_2^2)$ contributions to these terms for completeness, here 
we list only the
$O(u^2)$ terms, with a scalar field $F$ receiving 
the D-term contribution
of $u^2 g^2_\alpha Q_\alpha (F) Q_\alpha (S)$ in the square of its mass:
\begin{eqnarray} D_{16(10)} &=& g^2_\alpha (\cos\alpha ) u^2 \left( 
{1\over 6}\cos\alpha +
{1\over {2\sqrt{15}}}\sin\alpha\right)\, ,\\ D_{16({\bar 5})} &=&g^2_\alpha 
(\cos\alpha ) u^2
\left( {1\over 6}\cos\alpha - {3\over
{2\sqrt{15}}}\sin\alpha\right)\, ,\\ D_{16(1)} &=&g^2_\alpha (\cos\alpha ) 
u^2 \left( {1\over
6}\cos\alpha + \sqrt{5\over 12}\sin\alpha\right)\, ,\\ D_{10(5)} &=& 
g^2_\alpha (\cos\alpha )
u^2 \left( -{1\over 3}\cos\alpha - {1\over \sqrt{15}}\sin\alpha\right)\,
 ,\\ D_{10({\bar
5})} &=&g^2_\alpha (\cos\alpha ) u^2 \left( -{1\over 3}\cos\alpha + {1\over
\sqrt{15}}\sin\alpha\right)\, ,
\end{eqnarray} 
where the notation refers to the decomposition of E$_6$ 
under SO(10)(SU(5)) as
$27\rightarrow 16(10+{\bar 5}+1)+10(5+{\bar 5})+1(1)$ and the 16 
contains all of the
MSSM quark and lepton fields as well as $\nu^c$. In our
calculations, we assume 
$g_\alpha = g_Y$, where $g_Y$ and $g_\alpha$ are both normalized in the 
grand unified
group.  This is a very good approximation because $b_\alpha \approx b_Y$ 
for the one-loop beta functions.

There are a few points to be noted here. First, the D-terms induced by the 
singlet VEV
$u$ can be much larger that the electroweak ones. This can cause large 
shifts in the scalar
masses, in the positive or negative direction, depending on the sign of 
$\alpha$. As we
shall see in the next section, this implies nontrivial additional 
constraints on such a
scenario. Secondly, the presence of three complete {\bf 27}'s slows 
down the evolution of
the strong coupling $\alpha_s$. In fact at the one-loop level, it does 
not evolve at all
between the electroweak and GUT scales. As a result, the strong coupling 
stays stronger
than in the MSSM case as the scale increases to the GUT scale. Thus the 
gluino loop
contribution to the running masses turns out to be considerably higher 
than that in the
MSSM. And finally, there is a new, but small, RGE contribution from the 
$U(1)_\alpha$
gaugino loop.

\section{CONSTRAINTS FROM THE SCALAR MASS EXPRESSIONS}

As can be clearly seen from the previous section, there are several 
parameters ($m_0,
m_{\tilde{g}}, u$, $\tan \beta, A_0,m_t,\alpha_s$) determining the 
actual values of the
scalar masses. However, certain qualitative features are seen among 
the mass terms derived
with an additional broken U(1), which restrict the model considerably. 
We shall examine
these constraints below.

The $Z'$ mass is proportional to the singlet VEV $u$ for any given value 
of $\alpha$. The
experimental lower limit on $m_{Z'}$ is model-dependent. For the 
$\eta$-model it is about
440 GeV \cite{god}, while a model-independent analysis of the Tevatron data, 
assuming the $Z'$ to have
the same coupling strength as the $Z$, sets a limit of nearly 700 GeV 
\cite{egr}.  Setting $g_\psi$ =
$g_\chi$ = $g_Y$ leads to a relatively weaker coupling for the $Z'$, 
since it has no SU(2)
coupling except through mixing. Thus in a general anaylsis it is safe 
to assume that the
$Z'$ mass must be at least 500 GeV. This immediately translates to a 
minimum value of $u$
for any $\alpha$.

The higher $u$ is, the magnitudes of the D-terms are also 
correspondingly higher. The
most important consequence of this will be the way the parameters 
in the Higgs sector are
affected.  A desirable (although technically not absolutely required) 
condition to be satisfied for 
electroweak symmetry
breaking is 
\begin{eqnarray} m^2_{H_2} + {\mu}^2 < 0\, , 
\end{eqnarray} where the
Higgsino mass parameter $\mu$ is given at tree level by 
\begin{equation} {\mu}^2 = -
M^2_Z/2 + (m^2_{H_1} - m^2_{H_2} \tan^2 \beta)/(\tan^2 \beta - 1)
\end{equation} Also, there is the condition regarding the minimum 
value of the square of
the pseudoscalar mass:
\begin{equation} m^2_A = m^2_{H_1} + m^2_{H_2} + 2 {\mu}^2 > 0 
\, 
\end{equation} at tree level.  In the above, we note that the experimental 
bound on $m_A$ comes from the nonobservation of the decay $Z \rightarrow 
h + A$ which depends also on $m_h$.  In the limit of large $\tan \beta$, 
$m_h = m_A$, hence $m_A > M_Z/2$ may be used.  However, we will be 
conservative and use zero instead.  Also, the existence of the extra U(1) 
at the TeV scale changes the two-Higgs-doublet structure at the electroweak 
scale.  We have assumed these changes to be small enough so that the MSSM 
remains a valid approximation.  We now rewrite Eqs.~(18) and (20) as 
\begin{equation} -\frac{M^2_Z}{2} + \frac{\Delta m^2}{\tan^2 
\beta -1} < 0 \end{equation}
and
\begin{equation} -M^2_Z - \frac{\Delta m^2}{\cos 2 \beta} > 0 
\, ,
\end{equation} where
\begin{equation}
\Delta m^2 = D_{10({\bar 5})} - D_{10(5)} + \frac{3}{2} I\, . \end{equation}

Figs.~1 and 2 show some sample  parameter space of the $N$- and $\eta$-models,
respectively, to indicate the types of constraints implied by 
these inequalities. Clearly,
with a positive value of $\alpha$ it is rather difficult to 
satisfy the first condition
(e.g. Fig. 2(a) for the $\eta$-model), while still having $u$ 
large enough to be compatible
with the experimental limits on
$m_{Z'}$. The fact that $I$ is always positive makes the constraint 
even stronger. The
general expression for $I$ is linear in both $m^2_0$ and $m^2_{\tilde{g}}$. 
In order that
the damaging effect of $I$ is less, one requires $m_0$ (as also the gaugino 
mass) to be on
the lower side. Also, the large D-terms can be balanced only if $\tan\beta$ 
is sufficiently
large. To satisfy the second inequality, however, a low value of $m_0$ or
$m_{\tilde{g}}$ makes matters worse, as can be seen from Fig. 2(b).  
Note also that we need $\cos 2\beta$ here to be negative, 
hence $\tan \beta > 1$ is required. It turns out that in the 
$\eta$-model, the only
way to satisfy the conditions is to have $\tan \beta \approx$ 8 or 
higher. Models with
negative values of $\alpha$, like the $N$-model, are less tightly 
restricted because the
D-term constributions drive the first inequality in the right 
direction, favoring a large
$u$. However, the pseudoscalar mass limit prevents $u$ from being 
too large in this case
also. On the whole, in such models it is possible to have a smaller 
value of $\tan\beta$
than, say, in the $\eta$-model, as the curves in Fig. 1(a) and 1(b) indicate.

A further constraint on the $\eta$-model comes from the nature of the 
two types of D-terms
that contribute to the sfermion masses. Of them, $D_{16(10)}$ is always 
positive.
$D_{16({\bar 5})}$, however, can be negative for a positive value of 
$\alpha$. This in turn
implies large negative shifts in
$m^2_{\tilde{d}^c}$, $m^2_{\tilde{e}}$, and $m^2_{\tilde \nu}$. 
Of these,
$m^2_{\tilde{d}^c}$ receives a substantial positive RGE contribution from 
the gluino
loops. However, constraining the  ``left-handed" sneutrino masses  to be 
above 45 GeV
(the limits from LEP-1) requires  the universal scalar mass $m_0$ to be 
appropriately large,
or requires the universal gaugino mass $m_{1/2}=m_{\tilde{g}}$ to be big 
enough to produce
a compensating effect from the SU(2) and U(1) gaugino loops. This subjects the
$\eta$-model to an added restriction, as is illustrated in Fig. 3. In 
the $N$-model, on the
other hand, both
$D_{16({\bar 5})}$ and $D_{16(10)}$ are positive, and thus this model 
is free from such a
restriction. Another implication of this constraint is that the
$\eta$-model, or any model with a positive $\alpha$, is compatible with 
a no-scale
supergravity scenario (where
$m_0 = 0$) only if the universal gaugino mass is sufficiently high (at 
least about 300
GeV). We shall also see in the next section that the large positive values 
of $D_{16(10)}$, as well as the necessity of a large $m_0$, makes it difficult 
to have a light stop in this
scenario.

\section{SOME NUMERICAL RESULTS}

In our calculations we have assumed the GUT scale to be $2 \times 10^{16}$ 
GeV. The beta
functions for the different couplings are determined using \begin{eqnarray} 
b_1 = 3(3) +
\frac{3}{5}, ~~~ b_2 = -6 + 3(3) + 1,\\ b_3 = -9 + 3(3), ~~~ b_N = 3(3) + 
\frac{2}{5}\, .
\end{eqnarray} Also, we use $\alpha_{s} = 0.123$, $\alpha = 1/127.9$, and
$\sin^2 \theta_W = 0.2317$ at the scale $M_Z$. We take the top mass to be 
175 GeV.

The graphs presented in Fig. 4 show the masses of the sfermions belonging 
to the first two
generations plotted against the $Z'$ mass for some sample  parameter space. 
The $\eta$- 
and $N$-models are chosen to demostrate general features in the case of  
positive and negative $\alpha$ respectively. The 
values for the
other parameters chosen for the purpose correspond to regions in the 
parameter space which
satisfy the constraints discussed in the previous section.

In both of the sample scenarios, the squark masses are predicted to lie 
approximately in
the range 300-550 GeV, and the slepton masses in the range 100-450 GeV. 
This puts them
almost beyond the search limits of the upgraded Tevatron, and above 
the limits of
LEP-2. If they are discovered with lower masses, then that will render 
an additional U(1)
factor somewhat unlikely, since that would make it extremely difficult to 
satisfy the
constraints mentioned above, for a generic $\alpha$.  On the
other hand, the allowed range of the sfermion masses falls well within the 
discovery limits
of the LHC. It should be mentioned here that if SUSY has to be broken at a
scale not exceeding a TeV or so, then any additional $Z'$ is quite likely to 
be discovered at the LHC together with the sfermions.

On the other hand, Fig. 5 shows that the exotic (SU(2) singlet) squarks can 
have relatively
low masses. This is because $h^c$ has both $Q_\psi$ and $Q_\chi$ negative, 
while $h$ has a
positive $Q_\chi$ but an equally negative $Q_\psi$. As a result, the D-terms 
for them can
reduce the scalar masses. Thus in such a scenario the exotic squarks, which 
should be
produced on par with the ordinary ones at hadronic colliders, could be 
discovered {\it
before} the ordinary ones. Of course, it should be remembered all along 
that there is a
further uncertainty in the prediction of the exotic squark masses, since 
their F-term masses
are not predicted by the model. In fact, if the
$Z'$ mass is large then it is imperative for the corresponding exotic quarks 
to be heavy so that one does
not end up with an inadmissibly small value for $m^2_{\tilde{h}}$. 
Also, at least one of the three exotic quarks should be 
heavy for it to induce a sizable negative contribution by its Yukawa coupling 
to the square of the 
mass of $\tilde S$ to precipitate the spontaneous breaking of $U(1)_\alpha$ 
in the first place.

Finally, let us consider the stop masses in this framework. By virtue of the 
potentially large
${\tilde{t}}_L$-${\tilde{t}}_R$ terms arising through trilinear couplings in 
the third
generation, the stop mass matrix is given by \begin{equation} {\cal M}_{\rm 
stop} =\left(
\matrix{ m_{\tilde{Q}_3}^2 + m_t^2+ D_{16(10)}&m_t (A_t +
\mu
\cot\beta)\cr m_t (A_t + \mu \cot\beta)&m_{\tilde{t}^c}^2 + m_t^2+ D_{16(10)}
\cr}\right)\, ,
\end{equation} where $A_t$ is the trilinear scalar coupling for the third 
generation and we
have neglected the $O(v_1^2,v_2^2)$ D-term mass contributions in the 
expression. Diagonalization of this mass matrix implies that one physical 
state ($m_{\tilde t_1}$) can
be quite light  ($\leq 100$ GeV) if the off-diagonal 
element happens to be
large enough. This has attracted a lot of interest in recent times \cite{stop} 
because of two main
reasons, namely (1) a partial solution to the
$R_b$ problem is offered by a light stop scenario \cite{rb}, and (2) a light 
stop provides a better fit to $\alpha_{s}(m^2_Z)$ \cite{als}. Both of these 
require the stop to be lighter than,  or at most, about
100 GeV. It has been found that in the MSSM framework, a stop in this 
mass range can indeed
exist without contradicting the currently available data on top decay 
from the Tevatron,
although with accumulating luminosity one can restrict it further. While 
LEP-1 imposes a
lower limit of 45 GeV on the stop mass, the only other constraint on it 
is from a search
for the direct production of stop pairs, which closes a window in the 
range 65-88 GeV. Thus
a stop lighter than about 100 GeV is an object of active interest.

In our scenario, one unavoidable feature is the presence of the new D-term 
in the diagonal
mass terms $m^2_{\tilde t_{L,R}}$. This is the term $D_{16(10)}$ which is 
always positive.
The effect of such a term is to boost both the eigenvalues of the mass 
matrix. The only way
in which a light stop could still be envisioned is in cases where the 
off-diagonal terms are
also very large. This would require the magnitude of either $A_t$ or 
$\mu$ to be large
compared to the diagonal mass parameters. The former, however, is 
constrained by conditions
arising from color and charge invariance of the vacuum, as well as 
by flavor-changing
neutral current suppression \cite{[ccb]}. Our numerical estimates show that 
the lowest ($\approx 100$ GeV)
possible value of $m_{\tilde t_1}$ necessitates $A_0 \sim 1 ~{\rm TeV}$ 
where the above
conditions cannot be satisfied. On the other hand, a large magnitude of 
$\mu$ requires the universal scalar mass $m_0$ or $m_{\tilde{g}}$ to be 
correspondingly large, a
situation that would simultaneously increase the diagonal terms as 
well. Thus the general
conclusion is that a light stop of the type envisioned in the MSSM 
is rather unlikely when
there is an additional U(1) symmetry broken at such a scale that the 
additional gauge boson
has a phenomenologically permissible mass.

In Fig. 6 we show our expectations of the lighter stop mass in  regions of the
parameter space where $m_0$ and $m_{\tilde{g}}$ are close to their 
minimal possible
values. Although the
$N$-model allows smaller values for the lightest stop than the 
$\eta$-model does, due to possible smaller values of $m_0$ and 
$m_{\tilde{g}}$, even here we do not find a stop mass 
lighter than 120 GeV 
when we scan the viable parameter space.  The different predicted values of
$\mu$ and $f=|\mu /u|$ (the coupling of the superpotential term 
$SH_1H_2$ in the model) for the same parameter space 
as in Fig. 6 are
plotted in Fig. 7.

\section{SUMMARY AND CONCLUSIONS}

We have studied the scalar mass patterns predicted in a SUSY scenario 
embedded in a general
E$_6$ grand unified theory, with an extra U(1) gauge symmetry surviving 
at low energy. Our
observation is that, if one assumes scalar mass universality at the GUT 
scale, then the
conditions for electroweak symmetry breaking as well as a phenomenologically 
viable
pseudoscalar mass puts a general model under constraints. Such constraints 
are especially
tight for the $\eta$-model, as also for models where the angle $\alpha$ is 
negative. As far
as the standard sfermion masses are concerned, they are expected to be in a 
somewhat higher
range than in the MSSM case, although they can still be within the discovery 
limits of the
LHC. This scenario also allows the interesting possibility of lighter exotic 
squarks. And
lastly, it is difficult to accommodate a light stop in such a picture.

\section{ACKNOWLEDGEMENTS} B.M. acknowledges the hospitality of Ernest Ma 
and the
Department of Physics, University of California, Riverside, while this work 
was  in
progress. This work was supported in part by the U.~S.~Department of Energy 
under Grant No.~DE-FG03-94ER40837.

\newpage
\leftline{\Large \bf APPENDIX}
\vskip 0.5cm

In this paper we have discussed the case of a spontaneously broken U(1) gauge 
symmetry
together with the explicit soft breaking of supersymmetry. However, it is 
also
important to consider the possibility of preserving SUSY while breaking the
U(1) gauge symmetry, or more generally, while reducing the rank of the gauge 
symmetry, {\it e.g.} from SO(10) to SU(5). Suppose we want to break $G$ to 
$G'$ such that the rank of $G'$
is one less than the rank of $G$. Then there must be an U(1) gauge factor 
such that
\begin{equation} G \supset G' \times U(1),
\end{equation} and the scalar superfield which does it must transform under 
this $U(1)$. If
there is only one such superfield, it is clearly impossible to have a 
superpotential
invariant under $G$ and hence no spontaneous breaking of $G$ is possible 
without also breaking supersymmetry.

Consider now two scalar superfields, $\phi_1$ and $\phi_2$, transforming 
oppositely under
$G$ and hence also under $U(1)$, then the superpotential is \begin{equation} 
W = \mu \phi_1
\phi_2.
\end{equation} The supersymmetric scalar potential is then \begin{equation} 
V = |\mu
\phi_1|^2 + |\mu \phi_2|^2 + V_D, \end{equation} where $V_D \geq 0$ comes 
from the gauge
sector. The supersymmetric minimum of this $V$ clearly does not break the 
gauge symmetry. 

We now add a singlet superfield $\chi$, {\it i.e.} one which is trivial 
under $G$. The
superpotential is then
\begin{equation} W = \mu \phi_1 \phi_2 + f \phi_1 \phi_2 \chi + r \chi + 
{1 \over 2} M
\chi^2 + {1 \over 3} h \chi^3,
\end{equation} and the supersymmetric scalar potential becomes 
\begin{equation} V = |\mu
\phi_1 + f \phi_1 \chi|^2 + |\mu \phi_2 + f \phi_2 \chi|^2 + |f 
\phi_1 \phi_2 + r + M \chi
+ h \chi^2|^2 + V_D. \end{equation} A supersymmetric minimum of 
$V$ is obtained for
\begin{equation}
\langle \chi \rangle = - {\mu \over f}, ~~~ \langle \phi_1 \phi_2 
\rangle = - {r \over f} +
{{M \mu} \over f^2} - {{h \mu^2} \over f^3}, \end{equation} and $|
\langle \phi_1 \rangle| =
|\langle \phi_2 \rangle|$ from $V_D = 0$. 

Consider the possible effect on scalar masses due to the so-called 
D-terms. The latter are
proportional to $|\langle \phi_1 \rangle|^2 - |\langle \phi_2 
\rangle|^2$ and have thus no
effect as long as supersymmetry is maintained. In the presence of soft SUSY 
breaking by universal scalar masses at a chosen scale, the 
scalar-mass parameters at the symmetry breaking scale 
would differ by how they couple to the other particles of the theory in 
the evolution of
these couplings away from the chosen scale. In the minimal supersymmetric 
standard model
(MSSM), the two Higgs scalar doublets evolve differently because one couples 
to the $t$
quark (with its corresponding large coupling) and the other does not. 

Recently there have been discussions in the literature\cite{kmykm} regarding 
the contribution of D-terms to the masses of the MSSM sfermions from the 
reduction in rank of the gauge group at a very high scale.  This is of 
course possible, but in order to calculate the deviations for a particular 
pattern of symmetry breaking, an implicit assumption 
of universal scalar masses at some higher scale (for example the Planck 
scale $m_{pl}$) must be made.  Possible (but unknown) differences in the 
interactions of $\phi_1$ and $\phi_2$ with other fields between $m_{pl}$ 
and $\langle \phi_{1,2} \rangle$ would then generate these extra D-terms 
whose magnitudes are however not guaranteed to be observable. 
In this paper, our D-terms are directly related to the mass of the $Z'$ 
just as the usual electroweak D-terms are related to $M_Z$, 
and are thus subject to direct experimental verification.

\newpage
\leftline{{\Large\bf Figure captions}}
\begin{itemize}

\item[Fig. 1~:]{{(a): $(m_{H_2}^2 + \mu^2 )/ {\rm GeV}^2$ vs. $M_{Z'}/$GeV 
for $\alpha = N$
    with $m_{\tilde{g}} = 200$ GeV, $A_0=0$, and $\tan \beta =4$.\\
(b):  $m_A^2/{\rm GeV}^2$ vs. $M_{Z'}/$GeV for $\alpha = N$
    with $m_{\tilde{g}} = 200$ GeV, $A_0=0$, and 
$\tan\beta =4$.\\
In both case, the curves are labeled by the number $m_0/$GeV.}\label{fig1}}

\item[Fig. 2~:]{{(a): $(m_{H_2}^2 + \mu^2)/{\rm GeV}^2$ vs. $M_{Z'}/$GeV for 
$\alpha = \eta$ with $A_0=0$, and $\tan \beta = 8$. \\(b):  $m_A^2/
{\rm GeV}^2$ vs. $M_{Z'}/$GeV for $\alpha = \eta$ with $A_0=0$, and  
$\tan \beta = 8$.\\ In both cases, the curves are labeled by  the pair of 
numbers  
$\left( m_0/{\rm GeV},m_{\tilde{g}}/{\rm GeV} 
\right)$.}\label{fig2}}

\item[Fig. 3~:]{{Lower bound on $m_0/$GeV vs. $M_{Z'}/$GeV for $\alpha = \eta$
    from $m_{\tilde \nu_{e_L}} > 45$ GeV. The different curves represent
gluino masses from 150 GeV to 400 GeV at 50 GeV intervals.}\label{fig3}}

\item[Fig. 4~:]{{(a): $m_{\tilde f}/$GeV vs. $M_{Z'}/$GeV for $\alpha = N$
    with $m_{\tilde{g}} = 200$ GeV, $m_0=0$ GeV, $A_0=0$, and $\tan\beta
=4$. In descending order on the left-hand end of the graph, the curves
represent the masses of $\tilde{d_R}$, $\tilde{d_L}$, $\tilde{u_L}$,
$\tilde{u_R}$, $\tilde{e_L}$, $\tilde{\nu_{e_L}}$, and $\tilde{e_R}$. 
\\ (b):
$m_{\tilde f}/$GeV vs.
$M_{Z'}/$GeV for
$\alpha =
\eta$
    with $m_{\tilde{g}} = 200$ GeV, $m_0=200$ GeV, $A_0=0$, and 
$\tan\beta =8$. In  descending order on the left-hand end of the graph, the
curves represent the masses of $\tilde{d_L}$, $\tilde{u_L}$, $\tilde{u_R}$,
$\tilde{d_R}$, $\tilde{e_R}$,  $\tilde{e_L}$, and
$\tilde{\nu_{e_L}}$.}\label{fig4}}

\item[Fig. 5~:]{{(a): (exotic sfermion masses)$/$GeV vs. $M_{Z'}/$GeV 
for $\alpha = N$
    with $m_{\tilde{g}} = 200$ GeV, $m_0=0$ GeV, $A_0=0$, 
$\tan\beta =4$, and
$m_h=m_{h^c}=400$ GeV.\\ (b): (exotic sfermion masses)$/$GeV vs. 
$M_{Z'}/$GeV for $\alpha =
\eta$ 
    with $m_{\tilde{g}} = 180$ GeV, $m_0=200$ GeV, $A_0=0$, 
$\tan\beta =8$, and
$m_h=m_{h^c}=200$ GeV.}\label{fig5}}

\item[Fig. 6~:]{{(a): $m_{\tilde{t}_1}/$GeV vs. $M_{Z'}/$GeV for 
$\alpha = N$
    with $m_{\tilde{g}} = 200$ GeV, $m_0=0$ GeV, $A_0=0$, and 
$\tan\beta =4$.\\
(b): $m_{\tilde {t}_1}/$GeV vs. $M_{Z'}/$GeV for $\alpha = \eta$
    with $m_{\tilde{g}} = 200$ GeV, $m_0=200$ GeV, $A_0=0$, and 
$\tan\beta =8$.}\label{fig6}}

\item[Fig. 7~:]{{(a): $|\mu |/$GeV vs. $M_{Z'}/$GeV. 
The $\alpha =N$ case has $m_{\tilde{g}} = 200$ GeV, $m_0=0$ 
GeV, $A_0=0$, and $\tan\beta
=4$. The $\alpha =\eta$ case has $m_{\tilde{g}} = 200$ GeV, 
$m_0=200$ GeV, $A_0=0$, and
$\tan\beta =8$.\\ (b): $f=| \mu /u|$ vs. $M_{Z'}/$GeV. 
The $\alpha =N$ case has $m_{\tilde{g}} = 200$ GeV, 
$m_0=0$ GeV, $A_0=0$, and $\tan\beta
=4$. The $\alpha =\eta$ case has $m_{\tilde{g}} = 200$ GeV, 
$m_0=200$ GeV, $A_0=0$, and
$\tan\beta =8$.}\label{fig7}}

\end{itemize}
\end{document}